\documentclass[prl,showpacs,floatfix,twocolumn]{revtex4}
\usepackage{graphicx}
\newcommand{\be}{\begin{eqnarray}}
\newcommand{\ee}{\end{eqnarray}}
\def\beq{\begin{equation}}
\def\eeq{\end{equation}}

\begin{document}
\title{Universality in quantum chaos and the one parameter scaling theory}
\author{Antonio M. Garc\'{\i}a-Garc\'{\i}a}
\affiliation{Physics Department, Princeton University, Princeton,
New Jersey 08544, USA}
\affiliation{The Abdus Salam International Centre for Theoretical
Physics, P.O.B. 586, 34100 Trieste, Italy}
\author{Jiao Wang}
\affiliation{Temasek Laboratories, National University of
Singapore,117542 Singapore} \affiliation{Beijing-Hong
Kong-Singapore Joint Center for Nonlinear and Complex Systems
(Singapore), National University of Singapore, 117542 Singapore}
\begin{abstract}

We adapt the one parameter scaling theory (OPT) to the context of
quantum chaos. As a result we propose a more precise
characterization of the universality classes associated to
Wigner-Dyson and Poisson statistics which takes into account
Anderson localization effects. Based also on the OPT we predict a
new universality class in quantum chaos related to the
metal-insulator transition and provide several examples. In low
dimensions it is characterized by classical superdiffusion or a
fractal spectrum, in higher dimensions it can also have a purely
quantum origin as in the case of disordered systems. Our findings
open the possibility of studying the metal insulator transition
experimentally in a much broader type of systems.
\end{abstract}

\maketitle

Technological advances in recent years in the control of quantum
coherence have supposed an important stimulus to the field of
`quantum chaos', namely, the study of the effect of classical
chaos in quantum mechanics. A cornerstone result in quantum chaos
is the so called Bohigas-Giannoni-Schmit (BGS) conjecture
\cite{oriol} that  states quantum spectra of classically chaotic
systems are universally correlated according to the Wigner-Dyson
random matrix ensembles (WD) \cite{mehta} for scales comparable to
the mean level spacing, $ \Delta$. Similarly, according to the
Berry-Tabor-Gutzwiller conjecture \cite{tabor},  the level
statistics of systems whose classical counterpart is integrable
are well described by Poisson statistics, namely, the spectrum is
uncorrelated.

Despite its unquestionable success, these conjectures are not
always verified. Exceptions include the Harper model
\cite{harper}, and the ubiquitous kicked rotor \cite{kick}. In the
latter, for short time scales, both quantum and classical motion
are diffusive in momentum space. However, quantum diffusion is
eventually arrested due to destructive interference effects that
localize eigenstates in momentum space. In this limit spectral
correlations are described by Poisson not by WD statistics. This
counter-intuitive feature is usually referred to as dynamical
localization \cite{kick, fishman}.

Deviations from the BGS conjecture are also expected  for
eigenvalue separations $\delta E \sim \hbar/t_E$ due to weak
localization effects \cite{larkin}. The typical scale  $t_E =
\lambda^{-1}|\log \hbar|$ is the Ehrenfest time with $\lambda$ the
classical Lyapunov exponent. However, these corrections, though
universal, do not really invalidate the BGS conjecture since
$\delta E \sim \hbar/t_E \gg \Delta$.

From the previous discussion it is clear that the main deviation
from the semiclassical picture contained in the BGS conjecture is
due to localization effects.  This fact rises the following
questions: Why in some quantum chaotic systems localization is
only weak but in others is strong enough to induce a
metal-insulator transition? Is it possible to give a more precise
relation between classical motion and quantum features? Are there
any other universality classes in the context of non-random
Hamiltonians? The aim of this paper is to address these questions
within the framework of the one parameter scaling theory
\cite{one} introduced originally in the context of disordered
systems \cite{anderson}.

A key concept in this theory is the dimensionless conductance $g$
introduced  by Thouless \cite{thouless2}. It is defined either as
i) the sensitivity of a given quantum spectrum to a change of
boundary conditions in units of the mean level spacing $\Delta
\propto 1/L^d$,  or ii) as $g = E_c/\Delta$ where, $E_c$, the
Thouless energy, is an energy scale related to the diffusion time
to cross the sample.  In the metallic limit $E_c = \hbar
D_{clas}/L^2$ ($D_{clas}$ is the classical diffusion constant) and
therefore $g \propto L^{d-2}$. On the other hand if the particle
is exponentially localized due to destructive interference, $g
\propto e^{-L/\xi}$ where $\xi$ is the localization length and $L$
is the system linear size.

The change of $g(L)$ with the system size, $\beta(g)=
\frac{\partial \log g(L)}{\partial \log L}$, provides information
about localization. For $L \to \infty$, $\beta(g) = d-2>0$ in a
metal (without quantum corrections) and $\beta(g)= \log(g) <0$ in
an insulator. The OPT is based on the following two simple
assumptions:  i) The $\beta(g)$ function is continuous and
monotonous; ii) The change in the conductance with the system size
only depends on the conductance itself.

With this simple input the OPT is able to predict three
universality classes in the $L \to \infty$ limit: For $d >2$ and
disorder sufficiently weak, $g \to \infty$. The system has
metallic like features: Eigenfunctions are delocalized in real
space, and the spectrum is correlated according to WD statistics.
For $d \leq 2 $, or $d > 2$ and strong disorder, $g \to 0$. The
system is an insulator:  Eigenfunctions are exponentially
localized and the spectrum is correlated according to Poisson
statistics. For $d> 2$, a metal-insulator transition takes place
at a certain density of impurities and energy. It is characterized
by a size independent dimensionless conductance $g = g_c$, namely,
$\beta(g_c)= 0$. Eigenstates at the transition are multifractals
and the spectral correlations are universal \cite{kravtsov}, but
different from WD and Poisson statistics. We now study how the OPT 
 can be adapted to the context of quantum chaos.

{\it One parameter scaling theory in quantum chaos.-}The first
problem is the very definition of the Thouless energy. In
disordered systems it is estimated after an average over many
disorder realizations.  In quantum chaos no such ensemble average
can be carried out. However, a Thouless energy can still be
defined provided: i) The ensemble average is replaced by an
average over initial conditions; ii) The classical phase space is
homogeneous with no islands of stability. This condition
guarantees that the Thouless energy does not depend on the initial
conditions chosen.

The localization problem in quantum chaos is defined in momentum
space, not in real space. This can be understood as follows: In a
classically integrable systems the number of conserved quantities
(canonical momenta)  is equal to the dimensionality of the system.
In quantum mechanics each of these canonical momenta becomes a
good quantum number which labels the state. An integrable system
is thus localized in momentum space in the sense that there exists
a basis of momentum eigenstates in which the Hamiltonian is
diagonal and consequently the spectrum is uncorrelated (Poisson
statistics). As classical symmetries are reduced,  the Hamiltonian
is no longer diagonal in any momentum basis and eventually a
transition from Poisson to WD statistics is expected.

After these clarifications we are ready to define $g$ in
non-random Hamiltonians.  As the classical dynamics is by no means
restricted to standard diffusion, we study the more general case  $\langle p^2 \rangle \sim t^\mu$ ($\langle \ldots
\rangle$ stands for average over initial conditions) with $\mu
>0$. The Thouless energy is given by $E_c \propto N^{-2/\mu}$ where
$N$ is the system size, namely,  the size of the basis of momenta
eigenstates in which we express the Hamiltonian. The mean level
spacing is in many cases given by $\Delta \sim 1/N^{d}$, but there
are important exceptions: i) for periodic potentials, the Bloch
theorem applies, the spectrum is continuous, and $\Delta = 0$; ii)
for systems whose eigenstates are exponentially localized, $\Delta
\neq 0$ even in the $N \to \infty$ limit; iii) in systems with a singular continuous spectrum the scaling with the
system size may be anomalous $\Delta \propto N^{-d/d_e}$ ($d_e \neq
1$). A precise
definition of $d_e$ may depend on the system in question. In the
Harper model \cite{harper},  $d_e \approx 1/2$ stands for the
Hausdorff dimension of the spectrum. We are now ready to define
the dimensionless conductance in quantum chaos. In cases i) and ii)
above, $g \to \infty$ (metal) and $g = 0$ (insulator) respectively.
In case iii) (including $d_e =1$),  \be \label{g} g(N) =
\frac{E_c}{\Delta} = N^{\gamma_{clas}}, ~~~~~ \gamma_{clas} =
\frac{d}{d_e} - \frac{2}{\mu}. \ee The running of $g$ is thus
described by $ \beta(g) = \frac{\partial \log g(L)}{\partial \log
L}= \gamma_{clas} =  \frac{d}{d_e} - \frac{2}{\mu}$. Under the
assumptions of the OPT and using Eq. (\ref{g}) we propose the
following alternative definition of universality class in quantum
chaos: i) If $\gamma_{clas} >  0$, eigenfunctions are delocalized
as in a metal and the spectral correlations are described by WD
statistics. ii) If $\gamma_{clas} < 0$, eigenfunctions are
localized as in a insulator and  the spectral correlations are
described by Poisson statistics. iii) If $\gamma_{clas} =  0$,
eigenfunctions are multifractal, a metal insulator transition
takes places and the spectral correlations are universally
described by critical statistics \cite{kravtsov,sko}.

Several remarks are in order: i) Universality is restricted  to
scales of the order of the mean level spacing; ii) If $d_e \neq
1$, or for a continuous spectrum $\Delta=0$, it may not be
possible to carry out a statistical analysis of the spectral
correlations. In these cases our classificatory scheme still holds
but the characterization of the system as a metal or an insulator
must rely directly on the eigenfunctions statistics or the
transport properties; iii) Quantum destructive interference may
modify $g$. For instance, all eigenstates are localized in a 2$d$
disordered system despite the fact that semiclassically $g \propto
L^{d-2}$ is constant as at the Anderson transition; iv) In order
to define the Thouless energy uniquely the moments of the classical distribution of probability
must be described by a single scale.
\begin{figure}
\hspace{-.2cm}
\includegraphics[width=7.8cm,height=5.4cm,clip]{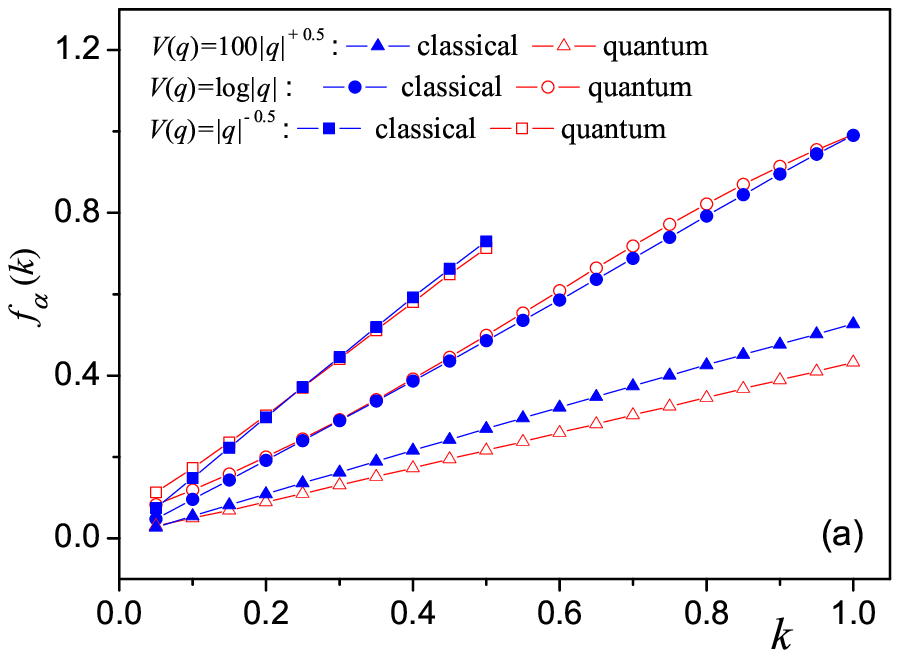}\vspace{-.5cm}
\includegraphics[width=7.8cm,height=5.4cm,clip]{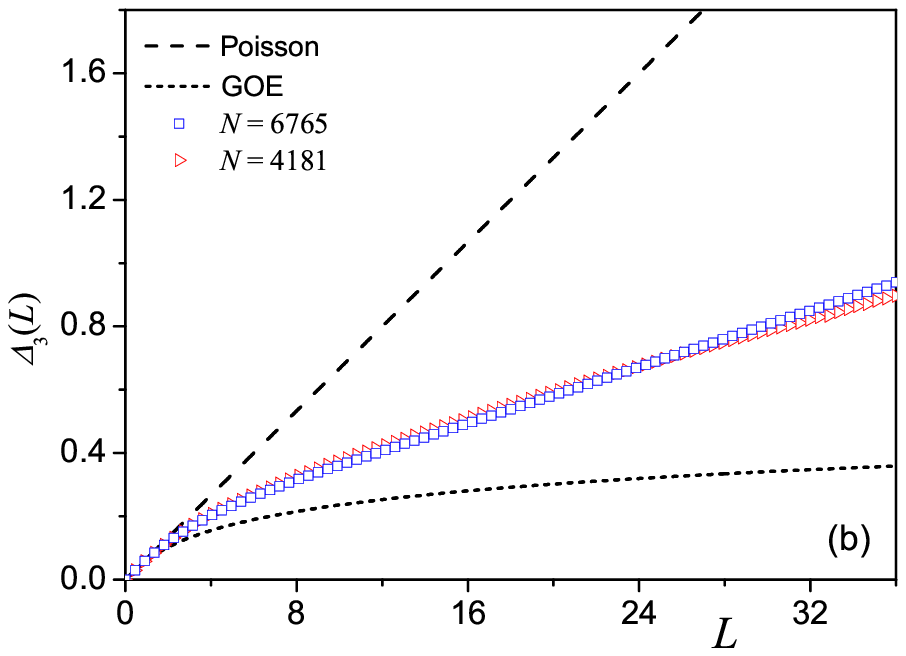}
\vspace{-.5cm} \caption{(Color online) (a) Comparison of the
classical and quantum $\langle |p|^{k} \rangle \propto t^{f_\alpha
(k)}$ for the case of a $1d$ kicked rotor with $V(q)\propto \log |q|, |q|^\alpha$. In agreement with the analytical
results the classical time dependence is not modified by quantum
corrections for $\alpha \leq 0$. (b) Spectral rigidity
$\Delta_3(L)$ for $V(q) = 10 \log|q|$. The level statistics in
this case have all the signatures of a metal insulator transition
such as scale invariance and linear $\Delta_3(L) \sim \chi L/15$ with $\chi
\approx 0.33 < 1 $.} \label{figure1}
\end{figure}

{\it Stability of the semiclassical predictions.-} The previous
discussion highlights the need of a clear understanding of the
effect of Anderson localization on $\beta(g)$. We focus on the $d=1$ and $d_e=1$ case,  the generalization $d > 1$ is straightforward. We investigate the
quantum dynamics of a classical diffusion process described by the
following fractional Fokker-Planck equation \cite{mirlin}, $
\left(\frac{\partial}{\partial t}- D_{clas}\frac{\partial^{2/\mu}}
{\partial |p|^{2/\mu}}\right)P(p,t)= \delta(p)\delta(t) \nonumber
$ where $D_{clas}$ is the classical diffusion coefficient and
$\mu$ is a real number. The moments of the distribution $P(p,t)$
(if well defined) are  $\langle |p|^k \rangle \propto t^{\mu
k/2}$, with $k$ a real number. The classical propagator in Fourier
space ($ t \leftrightarrow \omega$,  $p \leftrightarrow q$) is given by $K_0(q,\omega=0)=
\frac{\nu}{D_{clas}q^{{{2/\mu}}}} $ where $\nu$ stands for
the spectral density. One-loop
corrections due to interference effects \cite{mirlin} take the
form $K^{-1}(q)=K_0^{-1}(q)-\frac{(\pi\nu)^2}{2}
\int(dk)\frac{|q+k|^{2/\mu}-|k|^{2/\mu}} {|k|^{2/\mu}}$. The
quantum diffusion coefficient $D_{quant}$ to this order is easily
obtained by performing the integral above: \be
{D_{quant}}&=&D_{clas}-\mbox{C}  L^{{{{2/\mu}}}-1}   ~~~~ 1 <
{{{2/\mu}}} < 2 \nonumber \\
{D_{quant}}&=&D_{clas}-\mbox{C} \ln(qL) ~~~~~~~~~ {{{2/\mu}}} = 1\\
{D_{quant}}&=&D_{clas}-\mbox{C} q^{1-{{{2/\mu}}}}  ~~~~ 0
<{{{2/\mu}}} < 1 \nonumber \label{a3} \ee where $C$ is a different
constant for each case. The importance of the quantum effects
depends strongly on the value of ${{{2/\mu}}}$.

In the region $1 < {{{2/\mu}}} < 2$, quantum corrections diminish
the value of the classical diffusion constant. A renormalization
group analysis shows that the semiclassical prediction $\beta(g) <
0$ for $L \to \infty$ still holds but the transition to
localization is faster due to quantum corrections. In the region
$0<{{{2/\mu}}} <1$ quantum corrections are subleading with
respect to the classical term. As a consequence the semiclassical
prediction $\beta(g) > 0$ is not altered by quantum localization
effects. For ${{{2/\mu}}} = 1$,  the logarithmic behavior
resembles superficially that of 2$d$ disordered system,
$D_{quant}=D_{clas}-\mbox C\ln(L/l)$ where exponential
localization occurs. However there are differences. For anomalous
diffusion the mean free path, $l$ in the 2$d$ case, is the inverse
momentum $q^{-1}$. Therefore, the corrections to the bare coupling
constant are small for small momentum $q\sim 1/L$. Qualitatively
this implies the {\it absence} of eigenstate localization.  For a
rigorous proof based on the evaluation of $\beta(g)$ including
higher order terms we refer to \cite{mirlin} and references
therein. Therefore the quantum properties of an 1$d$ Hamiltonian
with classical dynamics such that $\langle p^k \rangle \propto
t^{k}$ resemble those of a disordered conductor at the
metal-insulator transition. In summary,  the semiclassical
predictions of a 1$d$ system whose classical dynamic is anomalous
are stable under quantum corrections.  In higher dimensions this
might change. For instance, in $d > 2$, we expect deviations due
to localization for $\gamma_{clas} > 0$.

{\it Examples. -} We now test the predictions of the OPT in
different  deterministic systems.\\
{\it 1$d$ Kicked rotor with classical singularities.} - We study
the quantum dynamics of, ${\cal H}= \frac{p^2}2
+V(q)\sum_n\delta(t -nT)$ for different non-analytic potentials
\cite{ant9,bao}, $V(q)=\epsilon |q|^\alpha$ and
$V(q)=\epsilon\log(|q|)$ with $q\in [-\pi,\pi)$, $\alpha \in
[-1,1]$ and $\epsilon$ a real number. The classical evolution is
dictated by the map
$p_{n+1}=p_{n}- \frac{\partial V(q_n)}{\partial q_n}$,
$q_{n+1}=q_n+Tp_{n+1}$ (mod$~2\pi$). In Ref. \cite{ant9} it was
found that for $\alpha> 0.5$ classical diffusion is normal.
Therefore $\langle p^2 \rangle \propto t$ and $\gamma_{clas} = -1
<0$. Thus the OPT predicts Poisson statistics even though the
classical dynamics is chaotic. It is remarkable that by using
scaling arguments dynamical localization can be predicted without
having to map the problem onto a 1$d$ Anderson model.

For $-0.5<\alpha < 0.5$, classical diffusion is anomalous,
$\langle p^k \rangle \sim t^{{k}(1-\alpha)}$. Using Eq. \ref{g},
\be \beta(g) = -\frac{\alpha}{1-\alpha}. \ee For $\alpha > 0 $ the
OPT predicts Poisson statistics no matter what the effect of
quantum corrections is. For $\alpha <  0$ we expect WD statistics
since quantum corrections do not modify qualitatively $\beta(g)$.
For $\alpha = 0$ ($\log$ singularity), we expect an Anderson
transition. Therefore dynamical localization can be overcome, even
in one dimension, if classical diffusion is fast enough; i.e. $\mu
> 2$.

These predictions have been tested by studying the quantum
evolution operator $\cal U$ over a period $T$ in a basis of plane
waves $|n\rangle$, $\langle m| {\cal U}| n \rangle =
\frac{1}{N}e^{-i2\pi M n^2/N} \sum_{l}e^{i\phi(l,m,n)}$ where
$\phi(l,m,n)= 2\pi (l+\theta_0)(m-n)/N-iV(2\pi (l+\theta_0)/N)$,
$l = -(N-1)/2,\ldots, (N-1)/2$ and $0 \le \theta_0 \le 1$;
$\theta_0 $ is a parameter depending on the boundary conditions
($\theta_0=0$ for periodic boundary conditions). The eigenvalues
and eigenvectors of $\cal U$ are computed by using standard
diagonalization techniques.
In Fig. 1a it is observed that, in agreement with the perturbative
analysis, the classical time dependence of $\langle p^k \rangle$
is not modified by quantum corrections for $\alpha \leq 0$. Thus a
genuine Anderson transition is expected for $\alpha =0$. The scale
invariance of the spectrum and the analysis of the level
statistics (see Fig. 1b) fully confirms the theoretical
prediction.

\begin{figure}
\hspace{-.2cm}
\includegraphics[width=7.8cm,height=5.4cm,clip]{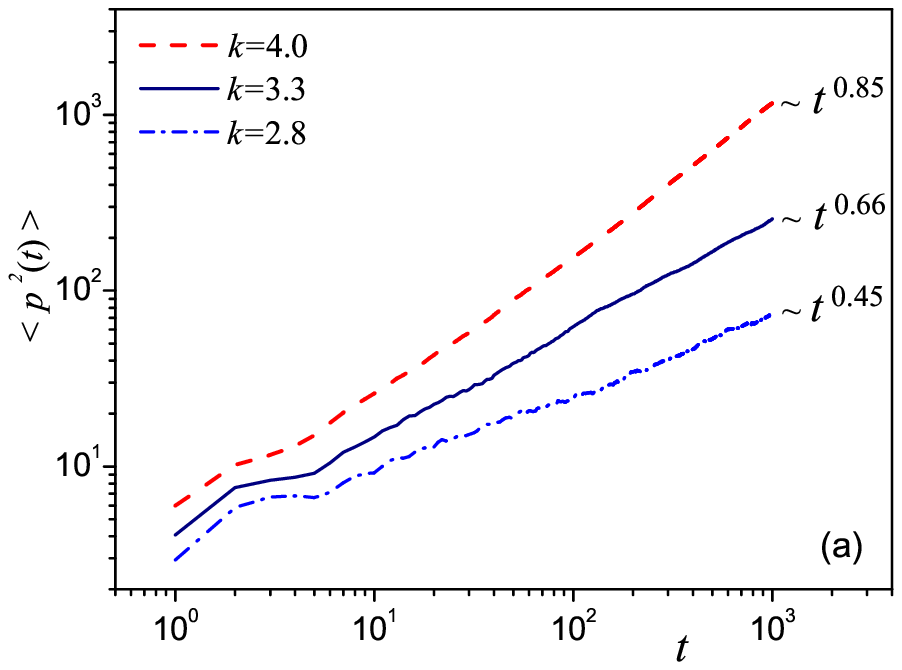}\vspace{-.5cm}
\includegraphics[width=7.8cm,height=5.4cm,clip]{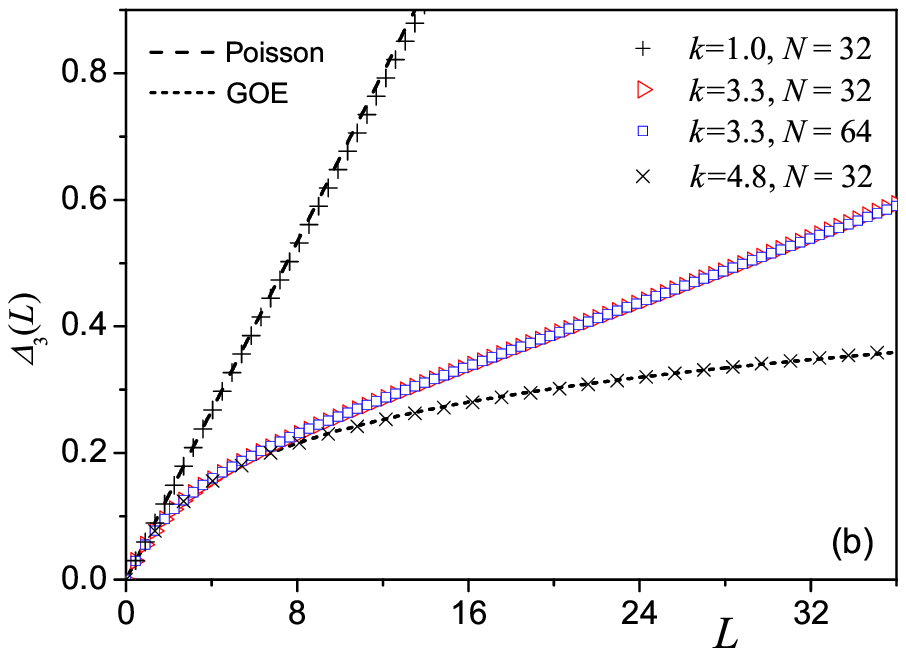}
\vspace{-.5cm}\caption{(Color online) (a) Quantum $\langle p^2
\rangle$ in the 3$d$ kicked rotor with a smooth potential for
three different kicking values: above, below and at the transition
$k = k_c \approx 3.3$. In agreement with the OPT $\langle p^2
\rangle \propto t^{2/3}$ at $k = k_c$. (b) The spectral rigidity
$\Delta_3(L)$ versus $L$ for the same kicking values. A transition
from WD to Poisson statistics is clearly observed as the kicking
strength decreases.  The level statistics at $k \sim k_c$ have all
the signatures of a 3$d$ Anderson transition such as scale
invariance and  $\Delta_3(L) \sim \chi L/15$ with $\chi \approx
0.19 < 1$. } \label{figure6}
\end{figure}

{\it 3$d$ Kicked rotor.-} We study the quantum dynamics of a 3$d$
 kicked rotor \cite{antwan7} with
a smooth potential (the 2$d$ version was studied in Ref. \cite{Doron}): ${\cal H}= \frac{1} 2 (\tau_1 p_1^2+\tau_2
p_2^2+\tau_3 p_3^2) +V(q_1,q_2,q_3)\sum_n\delta(t -nT)$ with
$V(q_1,q_2, q_3)=k\cos(q_1)\cos(q_2)\cos(q_3)$ and
$\tau_1,\tau_2,\tau_2$ incommensurate. 
The spectrum of the
evolution matrix was obtained by evolving a quantum state
$|\psi(0)\rangle$ and performing the Fourier transform of
$\langle\psi(t)|\psi(0)\rangle$. Classically the diffusion
$\langle p^2 \rangle \propto t$ is normal provided that the
classical phase space is fully chaotic. Quantum dynamics depends
strongly on $k$. In analogy with a 3$d$ disordered system, we
expect destructive interference stop the classical diffusion for
sufficiently small $k$. In the opposite limit, quantum effects are
small and diffusion persists. A careful finite size scaling
analysis \cite{sko} has confirmed this picture \cite{antwan7}. We
have found a metal-insulator transition at $k =k_c \approx 3.3$.
According to the OPT, since $\beta(g)=0$ at the transition,
quantum diffusion must be anomalous $\langle p^2 \rangle \propto
t^{2/3}$. Likewise, level statistics are described by WD (Poisson)
statistics in the limits $k  \gg (\ll) k_c$. For $k =k_c$ it is
expected spectral correlations be similar to those of a 3$d$
disordered system at the transition. As is shown in Fig. 2, the
numerical results fully agree with these theoretical predictions.

{\it Harper model.-} The 1$d$ Harper model \cite{harper}, ${\cal H} =
\cos(p) +\lambda \cos(2\pi \sigma x)$, with $\sigma$ irrational,
undergoes a metal insulator transition at $\lambda = 2$.
Classically the system is integrable.  However, the quantum motion
is diffusive $\langle x^2 \rangle \sim t^{2d_H}$ with $d_H$ the
Hausdorff spectral dimension.  With this information and Eq.
(\ref{g}) we can compute  the dimensionless conductance and $\beta
(g)$. As was expected  $\beta(g) = 0$  and $g =g_c$ is size
independent as it is expected at the metal insulator transition.
Our simple method predicts correctly the metal insulator
transition in this model as well.
In conclusion we have investigated under what conditions the one
parameter scaling theory can be utilized in quantum chaos. We have
utilized it to determine the number of universality classes in
quantum chaos and propose a more accurate definition of them. The
universality class related to the metal insulator transition has
been investigated in detail. We have tested our theoretical
predictions in different  kicked rotors and the Harper model. Our
findings open the possibility of studying the metal insulator
transition experimentally in a much broader class of systems.

AMG acknowledges financial support from a Marie Curie Outgoing
Action, contract MOIF-CT-2005-007300.  AMG thanks Igor Rozhkov
for illuminating discussions and a critical reading of the manuscript.
JW is grateful to Prof.
C.-H. Lai for his encouragement and support, and acknowledges
support from Defence Science and Technology Agency (DSTA) of
Singapore under agreement of POD0613356.

\end{document}